# Copyright and Creativity: Authors and Photographers

Douglas A. Galbi[1]
Federal Communications Commission

October 29, 2003

*Abstract*

The history of the occupations "author" and "photographer" provides an insightful perspective on copyright and creativity. The concept of the romantic author, associated with personal creative genius, gained prominence in the eighteenth century. However, in the U.S. in 1900 only about three thousand persons professed their occupation to be "author." Self-professed "photographers" were then about ten times as numerous as authors. Being a photographer was associated with manufacturing and depended only on mastering technical skills and making a living. Being an author, in contrast, was an elite status associated with science and literature. Across the twentieth century, the number of writers and authors grew much more rapidly than the number of photographers. The relative success of writers and authors in creating jobs seems to have depended not on differences in copyright or possibilities for self-production, but on greater occupational innovation. Creativity in organizing daily work is an important form of creativity.

**Note**: This paper is from Appendix C of *Sense in Communication*, available at www.galbithink.org

---

[1] The opinions and conclusions expressed in this paper are those of the author. They do not necessarily reflect the views of the Federal Communications Commission, its Commissioners, or any staff other than the author. I am grateful for numerous FCC colleagues who have helped me and encouraged me over the past eight years of my career at the FCC. Author's address: Douglas.Galbi@fcc.gov; FCC, 445 12'th St. SW, Washington, DC 20554, USA.

Voluminous discourse about authorship emerged in the eighteenth century as part of efforts to support writing as an occupation. Writers struggled to establish their personal genius, autonomous civic conscience, and economic viability while at the same time distancing themselves from the vulgar demands of public opinion and ordinary commerce.[1] This work, known in late twentieth century legal and literary scholarship as romantic authorship, was rather attractive to writers. By 1794, sixty works discussing authorship, written in German in the previous twenty years, were reviewed in a book entitled *Toward a Clarification of the Property and Property Rights of Writers and Publishers and of the Mutual Rights and Obligations. With Four Appendices. Including a Critical Inventory of All Separate Publications and of Essays in Periodical and Other Works in German Which Concern Matters of the Book as Such and Especially Reprinting.*[2] According to a leading contemporary scholar, in the eighteenth century:
> *writers set about redefining the nature of writing. Their reflections on this subject are what, by and large, gave the concept of authorship its modern form.*[3]

The resulting romantic concept of authorship still strongly shapes many authors' self-understanding of the work they do.[4]

Writing about authorship, not surprisingly, also encompassed writing laws. In England in 1710, the Statute of Anne established for authors and their assignees an exclusive, fourteen-year right to print books and other writings.[5] The statute's stated purpose was to promote the occupation of author – to prevent the economic harm that unauthorized printing was causing to authors and their families, and to encourage writing of useful books. In the U.S. in the 1780s, twelve out of the thirteen colonies passed copyright laws literally directed to securing author's rights.[6] The U.S. Constitution, written in 1787, included a clause empowering congress "[t]o promote the progress of science and useful arts, by securing for limited times, to authors and inventors, the exclusive right to their respective writings and discoveries."[7] The first U.S. Congress, meeting in 1790, promptly wrote and passed a federal copyright act.[8] In France, the opportunity to receive an exclusive right to print a work was extended to authors in a new royal regulation issued in 1777.[9] Four years after the French Revolution swept away royal power, the

---

[1] Over the past two decades scholars have done much work on the history of authorship and copyright. See, e.g. Woodmansee (1984), Boyle (1988), Ginsburg (1990), Jaszi (1991), Hesse (1991), Saunders (1992), Rose (1993), Woodmansee and Jaszi (1994), and Rice (1997). This literature has largely failed to explore over time how many persons actually made their living as authors.

[2] The title is from Ernst Martin Gräff's book, cited in Woodmansee (1984) p. 440. As a follow-up to relatively important work on footnotes, scholars not interested in pressing contemporary problems in communications policy might consider the history of titles, the style of which has changed significantly over time. For some preliminary research, see Darnton (1990) pp. 98-103.

[3] Woodmansee (1984) p. 426.

[4] Cf. Hepzibah's thoughts and actions in Hawthorne (1851) pp. 31-49, 241.

[5] 8 Anne c. 19. Facsimile and transcription available at http://www.copyrighthistory.com/anne.html

[6] Patterson (1968) pp. 181-92.

[7] Art I., § 8, cl. 8. On Aug. 18, 1787, James Madison and Charles Pinckney each submitted to the Constitutional Convention a list of additional, enumerated federal powers. Madison's list included, "To secure to literary authors their copyright for limited time." Pinckney's list included, "To secure to authors exclusive rights for a limited time." See Patterson (1968) pp. 192-3. The distinction between "literary authors" and "authors" is significant. In 1810, the French government established a separate regime of prepublication review for publications of twenty or fewer pages. More generally, early French literary law fostered the professionalization of writing and the transformation of the republic of letters into a republic of books. See Hesse (1991) pp. 245-6.

[8] 1 Stat. 124.

[9] Hesse (1991) p. 12.

French National Convention in 1794 passed a law protecting "property in the production of genius." The law granted to authors, heirs, and assignees exclusive publication rights for the lifetime of the author plus ten years.[1] All these laws were passed before industrial revolutions dramatically changed the nature of work.

Despite great changes in typical daily work since the eighteenth century, creativity is not usually associated with ordinary daily work. If labor (workers) and capital (machines) are considered to be substitutable inputs in a production process, human work is understood to be like that of machines. Machines might then be described as dead labor; that is, the essential value extracted from workers and re-embodied so that it can continue to work and never complain. In high-income societies today, the distribution of jobs has shifted greatly toward services; relationships among women, men, and children are much different; and information and communication technologies are much more powerful. Yet the eighteenth-century understanding of creativity remains. Creativity is some mysterious blessing from somewhere upon some persons, at some times. Creativity occurs in a realm separate from the work that keeps the lights on, the water running, the lawn neat and green, and puts food in the bowl for the cat and on the table for the children. Promoting creativity depends on a narrow legal field of reproductive rights historically centered on copyright.

The occupational history of photographers and authors in the U.S. suggests a much different understanding of creativity.[2] In 1850, only 82 males claimed the occupation of author in the first occupational census of the U.S.[3] In 1900, self-professed photographers were about ten times as numerous as self-professed authors. Being a photographer was associated with manufacturing and depended only on mastering technical skills and making a living. Being an author, in contrast, was an elite status associated with science and literature. Across the twentieth century, the number of writers and authors grew much more rapidly than the number of photographers.[4] The relative success of writers and authors in creating jobs seems to have depended not on differences in copyright or possibilities for self-production, but on greater occupational innovation. Creativity in organizing daily work is an important form of creativity.

---

[1] Id. pp. 118-21. As id. emphasizes, the law also abolished existing royal privileges for printing works and gave all citizens the right to print works formerly under royal privilege.

[2] The scholarly literature on communications tends to divide by sensory mode or media. This division obscures dimensions of communication central to communications industry growth. See Galbi (2003). Hawthorne (1851) recognizes commonalities in the occupations of writer and photographer (daguerreotypists) by having Holgrave both write stories and make daguerreotypes.

[3] In 1853, perhaps 420 new books, not reprinted or translated from foreign books, were published in the U.S. Tebbel (1972) v. II, p. 23, cf. id. v. I, p. 222. Reconciling this figure with the census figure for authors requires more scholarly work. Woodmansee (1984) p. 443, ft. 17 cites a "contemporary catalog of German writers" that estimated the number of writers in Germany in 1800 as 10,650. Id p. 431 also notes, "The professional writer emerged considerably later in Germany than in England and France." Reprinting of foreign works was prevalent in the U.S. in nineteenth century. See McGill (2003). About 1850, about 40% of new titles printed in the U.S. were reprintings of foreign works. See Tebbel (1972) v. II, p. 23; id. v. I, pp 221-2; cf. McGill (2003) p. 279, n. 2. Reprinting, however, cannot account for the huge difference in estimates of the number of authors in the U.S. and Germany. Rather than focusing on a few prominent authors, the history of authorship might focus more comprehensively on the number, background, and scope of activities of real persons involved in writing books.

[4] The number of authors (and closely related occupations) grew much faster than the overall population. Cf. Nunberg (1996) pp. 23-4.

## I. Bureaucratic Facts

| Year | Classifications in Hierarchy | Occupation | Number |
|---|---|---|---|
| | Table C1 | | |
| | **Classification and Number of Photographers and Related Jobs** | | |
| 1850 | none -- flat list | Daguerreotypists | 938 |
| 1860 | none -- flat list | Daguerreotypists | 2,650 |
| | none -- flat list | Photographers | 504 |
| 1870 | Manufactures, Mechanical and Mining Industries | Daguerreotypists and photographers | 7,558 |
| 1880 | Manufacturing, mechanical, and mining industries | Photographers | 9,990 |
| 1890 | Manufacturing and mechanical industries | Photographers | 20,040 |
| 1900 | Manufacturing and mechanical pursuits / miscellaneous industries | Photographers | 26,941 |
| 1910 | Professional service | Photographers | 31,775 |
| 1920 | Professional service | Photographers | 34,259 |
| 1930 | Professional service | Photographers | 39,529 |
| 1940 | Professional and Semiprofessional Workers / Semiprofessional Workers / Other semiprofessional workers | Photographers | 37,641 |
| 1950 | Professional, Technical | Photographers | 54,734 |
| 1960 | Professional, Technical, and Kindred Workers | Photographers | 45,393 |
| 1970 | Professional, Technical, and Kindred Workers / Writers, Artists and Entertainers | Photographers | 65,960 |
| 1980 | Managerial and professional specialty occupations / Professional speciality occupations / Writers, Artists, Entertainers, and Athletes | Photographers | 94,762 |
| 1990 | Managerial and professional specialty occupations / Professional speciality occupations / Writers, Artists, Entertainers, and Athletes | Photographers | 143,520 |
| 2000 | Arts, Design, Entertainment, Sports, and Media Occupations /Media and Communication Equipment Workers / | Photographers | 131,000 |
| | Arts, Design, Entertainment, Sports, and Media Occupations /Media and Communication Equipment Workers / | Television, video, and motion picture camera operators and editors | 43,000 |

Early commercial practice of photography rapidly and enduringly established "photographers" as an occupational category in government statistics. In 1850, the first U.S. occupational census included "daguerreotypists" as an occupational category. "Daguerreotypists" describes persons using the first widely known photographic process. This process was primarily used in the commercial provision of personal portraits. Other photographic processes developed rapidly. The evolution of nomenclature from "daguerreotypists" to the more general term "photographers" in censuses from 1860 to 1880 reflects this development. All subsequent censuses through to the 2000 census include the occupational category "photographers." The 2000 census also included a new category, "Television, video, and motion picture camera operators and editors." The introduction of this new category is an unusual development. The reduction in the technical complexity of making photographs, the development of new image-making technology, and the emergence of new fields, such as product design, advertising, and public relations, has not led to any other new occupational categories classified closely with photographers. What photographers do has been commonly and distinctively understood since the nineteenth century, and it has changed little since then.

The position of photographers in the job classification schema has, however, changed significantly from 1850 to 2000. Through the 1900 census, photographers were classified under manufacturing and mechanical industries. In 1910, photographers moved into the class "professional service." In 1940, a sub-heading "semiprofessional workers" was established, and photographers were placed under it. In 1950, that subheading was eliminated, and the higher level heading was changed from "Professional and semiprofessional workers" to "Professional, technical." These classification changes suggest struggles over the boundary of "professional." By 1980, photographers were being placed under a subheading "writers, artists, entertainers, and athletes." In 2000 "photographers" came under the subheading "media and communication *equipment* workers" [italics added]. Thus in the 2000 occupational classification, and in 1900 and earlier ones, photographers have been associated with machine workers.

| Table C2 Classification and Number of Authors and Related Jobs ||||
|---|---|---|---|
| Year | Classifications in Hierarchy | Occupation | Number |
| 1850 | none -- flat list | Authors | 82 |
|      | none -- flat list | Editors | 1,372 |
|      | none -- flat list | Reporters | 138 |
| 1860 | none -- flat list | Authors | 216 |
|      | none -- flat list | Editors | 2,994 |
|      | none -- flat list | Reporters | 411 |
| 1870 | Professional and personal services | Authors and lecturers | 458 |
|      | Professional and personal services | Journalists | 5,286 |
| 1880 | Professional and personal services | Authors, lecturers, and literary persons | 1,131 |
|      | Professional and personal services | Journalists | 12,308 |
| 1890 | Professional service | Authors and literary and scientific persons | 6,714 |
|      | Professional service | Journalists | 21,849 |
| 1900 | Professional service / Literary and scientific persons | Authors and scientists | 5,817 |
|      | Professional service | Journalists | 30,038 |
| 1910 | Professional service / Authors, editors, and reporters | Authors | 4,368 |
|      | Professional service / Authors, editors, and reporters | Editors and reporters | 34,382 |
| 1920 | Professional service / Authors, editors, and reporters | Authors | 6,668 |
|      | Professional service / Authors, editors, and reporters | Editors and reporters | 34,197 |
| 1930 | Professional service / Authors, editors, and reporters | Authors | 12,449 |
|      | Professional service / Authors, editors, and reporters | Editors and reporters | 51,844 |
| 1940 | Professional and Semiprofessional Workers / Professional Workers / Authors, editors, and reporters | Authors | 14,126 |
|      | Professional and Semiprofessional Workers / Professional Workers | Editors and reporters | 63,493 |
| 1950 | Professional, Technical | Authors | 16,184 |
|      | Professional, Technical | Editors and reporters | 91,472 |

| Year | Category | Occupation | Count |
|---|---|---|---|
| 1960 | Professional, Technical, and Kindred Workers | Authors | 20,734 |
| | Professional, Technical, and Kindred Workers | Editors and reporters | 63,279 |
| 1970 | Professional, Technical, and Kindred Workers / Writers, Artists and Entertainers | Authors | 26,004 |
| | Professional, Technical, and Kindred Workers / Writers, Artists and Entertainers | Editors and reporters | 152,984 |
| 1980 | Managerial and professional specialty occupations / Professional speciality occupations / Writers, Artists, Entertainers, and Athletes | Authors | 45,748 |
| | Managerial and professional specialty occupations / Professional speciality occupations / Writers, Artists, Entertainers, and Athletes | Editors and reporters | 210,831 |
| | Managerial and professional specialty occupations / Professional speciality occupations / Writers, Artists, Entertainers, and Athletes | Technical writers | 49,596 |
| 1990 | Managerial and professional specialty occupations / Professional speciality occupations / Writers, Artists, Entertainers, and Athletes | Authors | 106,730 |
| | Managerial and professional specialty occupations / Professional speciality occupations / Writers, Artists, Entertainers, and Athletes | Editors and reporters | 266,543 |
| | Managerial and professional specialty occupations / Professional speciality occupations / Writers, Artists, Entertainers, and Athletes | Technical writers | 74,292 |
| 2000 | Arts, Design, Entertainment, Sports, and Media Occupations / Media and Communications Workers / Writers and Editors | Writers and authors | 126,000 |
| | Arts, Design, Entertainment, Sports, and Media Occupations / Media and Communications Workers | News analysts, reporters and correspondents | 78,000 |
| | Arts, Design, Entertainment, Sports, and Media Occupations / Media and Communications Workers / Writers and Editors | Technical writers | 57,000 |
| | Arts, Design, Entertainment, Sports, and Media Occupations / Media and Communications Workers / Writers and Editors | Editors | 122,000 |

Over the past hundred and fifty years, authors have become much less distinguished in occupational category and classification. Authors had their own occupational category in 1850 and 1860, while from 1870 to 1900, authors were variously categorized with lecturers, literary persons, and scientists. These elite groups have high barriers to entry. By 1980, authors had been placed within a higher-level category, "Writers, Artists, Entertainers, and Athletes." "Technical writers," a category with a large number of workers, was also introduced in 1980 as a category with the same classification as "authors." The 2000 census included this category, but changed the category "authors" to "writers and authors." These changes indicate innovation and diversification in jobs associated with authors.

In occupational categorization, reporters and editors have been consistently distinguished from authors and writers, but not from each other. In 1850 and 1860, reporters and editors were separate categories, but from 1870 to 1990 they were grouped in a single category as "journalists" or "editors and reporters." In 2000, reporters and editors were again placed in two different categories. Originality provides one criterion for attempting to distinguish among authors, writers, reporters, and editors, while authority provides another, different, one.

Nonetheless, editors sometimes effectively act as writers, and reporters as authors. Moreover, some authors attempt to act as their own editors, but might be regarded as not succeeding in doing so. In any case, authors, writers, reporters, and editors have been consistently grouped together in the next higher level of occupational classification.

**II. Law and Economics**

While photographers have always been separated from authors in occupational classification, photographers have been recognized as authors under copyright law. In the early 1860s, photographically reproduced photographic portraits of famous persons became an important item of commerce.[1] The U.S. Congress passed a copyright statute for photographers in 1865.[2] In *Burrow-Giles Lithographic Co. v. Sarony* (1884), the U.S. Supreme Court confirmed that Congress acted within the constitutional scope of federal authority in granting copyrights to photographers, understood as authors:

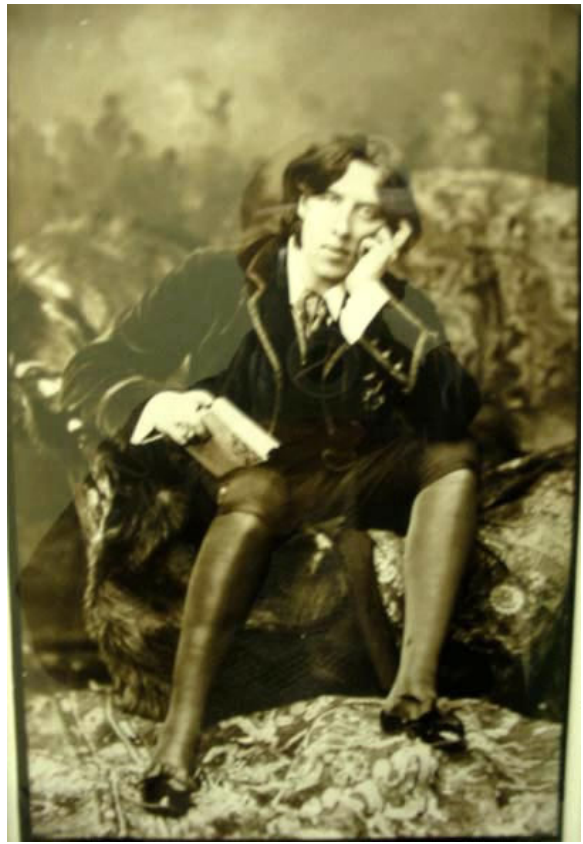

> *the constitution is broad enough to cover an act authorizing copyright of photographs, so far as they are representatives of original intellectual conceptions of the author. … These findings, we think, show this photograph* ["Oscar Wilde, No. 18"] *to be an original work of art, the product of plaintiff's intellectual invention of which plaintiff is the author, and of a class of inventions for which the constitution intended that congress should secure to him the exclusive right to use, publish, and sell…*[3]

The plaintiff in this case was Napoleon Sarony, who had been commercially and culturally prominent as a photographer for at least two decades.[4] Sarony specialized in celebrity photographs. From his studio on Broadway, he made photographs of most of the stars of New York theatre, including Sarah Barnhart, whom he paid $1500 for the opportunity.[5] Sarony himself did not generally operate the camera; he directed his efforts toward arranging the subject

---

[1] These objects were known as "cartes de visite." See Leggat (1999) and Stratford Hall Plantation (2002).
[2] Copyright Act, ch. 126, 13 Stat. 540-41 (1865).
[3] 111 U.S. 53, 59, 60.
[4] In the last third of the nineteenth century, Sarony was the best-known portrait photographer in the U.S. In 1897, a photographic journal lauded him as "the father of artistic photography in America." Sarony, "an inveterate joiner and a gadabout in New York City's literary and artistic circles," often wore a tasseled red fez. The reception room of his studio featured a stuffed crocodile hanging from the ceiling. Bassham (1978) pp. 4-6, 13, 16-7.
[5] Id. Sarony paid Lillie Langtry, called in her time "the world's most beautiful woman," $5000 for the exclusive right to make and sell portraits of her. For one such portrait, see id. p. 73. I'm not impressed. I think that many persons have known women more beautiful than this image.

and evoking from her or him the expression that he sought. He was an author of photographs in this way, and also as the owner of the studio that produced photographs.[1]

*Burrow-Giles* did not decide that all photographers were authors. In 1884, when *Burrow-Giles* was decided, persons describing their occupation as photographer were about ten times more numerous than persons describing their occupation as author. The situation and activities of most photographers were much different from those of Napoleon Sarony. Most photographers produced, on a commercial basis, conventional photographic portraits of ordinary persons that evoked among family and friends a sense of presence of the photographed persons. *Burrow-Giles* noted:

> *it is said that…a photograph is the mere mechanical reproduction of the physical features or outlines of some object, animate or inanimate, and involves no originality of thought or any novelty in the intellectual operation connected with its visible reproduction in shape of a picture. … This may be true in regard to the ordinary production of a photograph, and that in such a case copyright is no protection. On the question as thus stated we decide nothing.*[2]

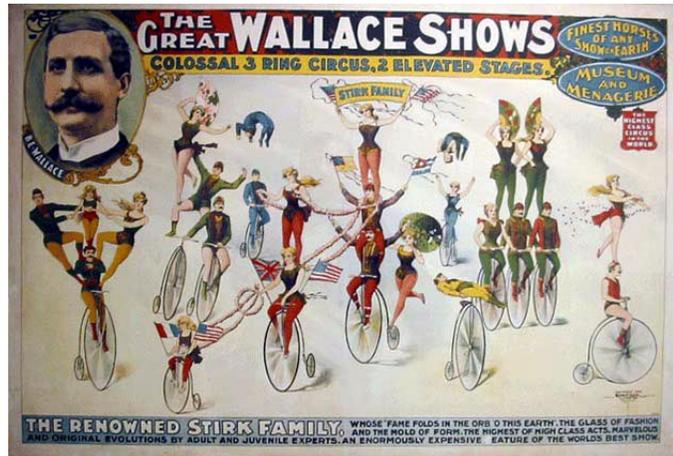

The court did not include in its officially published opinion a copy of Sarony's photograph. Including a copy probably would have advanced significantly appreciation for the opinion's legal distinction, at the cost of raising practical challenges and additional, important legal questions.[3] Deciding little in a way that limited communication of the decision was a pragmatic judicial strategy. The status of author, then and now, had little practical importance to most photographers' businesses and most persons' interest in photographs.

Questions of authorship can, however, implicate fundamental human concerns. In *Bleistein v. Donaldson Lithographing Co.* (1903), a newly elevated Supreme Court justice included these sentences in the Court's decision:

> *Others are free to copy the original. They are not free to copy the copy. The copy is the personal reaction of an individual upon nature. Personality always contains something unique. It expresses its singularity even in handwriting, and a very modest grade of art*

---

[1] Bassham (1978) pp. 14-5.
[2] 111 U.S. 53, 58-9.
[3] Id. at 55 mentions "the picture in suit, Exhibit A," but the official report of the case does not reproduce that picture. Photolithography was just becoming practical late in the nineteenth century, and photographs were not commonly reproduced in magazines until the twentieth century. If the photograph could have been printed in court reporters, doing so would have required advancing doctrines of use context as a limit on copyright. Legal disputes concerning the printing of judicial opinions have been important in shaping U.S. copyright law. See, e.g. Wheaton v. Peters, 33 U.S. 591 (1834). For a pioneering and endearingly unsophisticated discussion of U.S. judges' use of non-verbal material in printed opinions, see Dellinger (1997).

> *has in it something irreducible, which is one man's alone. That something he may copyright unless there is a restriction in the words of the act.*[1]

These sentences, forceful and impressive, proclaim the significance of an author in every persons' every act. Most persons, regardless of profession, would probably desire the elite status of author, under copyright or more generally. However, how all the effects of persons remain separate from nature and originals, so that a copy of a copy does not contain something personal and unique, "which is one man's alone," is unclear.[2] Moreover, it is clearly incorrect that copyright exists "unless there is a restriction in the words of the act." In the U.S., legal recognition of copyright depends on statutory provision of it.

The above sentences from *Bleistein* are probably best understood as invoking general appreciation for human creativity in a case that concerned posters advertising a circus. The original defendant described the circus advertisements as "tawdry pictures"[3] His counsel objected that "one, the Ballet, …is an immoral picture," and cited a case that invalidated copyright to a song because it used the word "hottest" in a way judged to be "indelicate and vulgar."[4] Abstracting from the vulgarities of particular images, the Court's decision argued that every person is special, a unique individual. The effects of every person are necessarily rare, valuable, and worthy of protection. In effect, every person essentially has the elite status of author.

That every person is an author, even just in ordinary work and family life, has been subjected to withering critical attack since at least the late nineteenth century. In a personal letter in 1909, a prominent U.S. public intellectual wrote:

> *[Henry James], like his brother and the parsons, attaches a kind of transcendental value to personality; whereas my bet is that we have not the kind of cosmic importance that the parsons and philosophers teach. I doubt if a shudder would go through the spheres if the whole ant heap were kerosened. Of course, for man, man is the most important theme.... Man of course has the significance of fact, but so has a grain of sand. I think the attitude*

---

[1] 188 U.S. 239, 249-50 (Feb. 2, 1903). Justice Oliver Wendell Holmes, who took his judicial oath for the Supreme Court on Dec. 8, 1902, delivered the opinion of the court. It was the second Supreme Court opinion that he wrote.

[2] The phrase "the personal reaction of an individual upon nature" might mean a person's reaction to nature, where that reaction might be fixed only in some realm separate from nature. Alternatively, "the personal reaction of an individual upon nature" might mean a person acting upon nature so as to augment it in a way uniquely attributable to that particular person. Neither interpretation seems consistent with the modern scientific view that human beings are part of created nature and continually transform the natural world. For discussion of related issues and interesting French case law, see Edelman (1994).

[3] 188 U.S. 241.

[4] Id. at 240, 247-48, citing Broder v. Zeno Mauvais Music Co. (1898), 88 F. 74. Broder v. Zeno concerned rights to a song that included the phrase "She's the hottest thing you ever seen." The Circuit Court judge's opinion noted: "I am of the opinion that the word 'hottest,' as used in the chorus of song 'Dora Dean,' has an indelicate and vulgar meaning, and that for that reason the song cannot be protected by copyright." Id. at 79. Considerations of immorality became relevant to copyright from early nineteenth-century English common law. See Wilkinson (1978). Neither the opinion nor the dissent in *Broder* makes reference to this issue, but it may well have been a significant judicial concern at that time. Concerns about indelicate, vulgar, or immoral material led in U.S. case law to a definition of obscene. However, obscenity is probably no longer valid legal grounds for defense against copyright infringement. See Mitchell Brothers Film Group v. Cinema Adult Theatre, 604 F.2d 852 (5'th Circ., 1979) and Nimmer on Copyright § 2.17. Nonetheless, the importance of morality to law and policy remains. For example, indecency in radio and television programming is a major concern of at least one current FCC commissioner. Indecent programming includes obscene programming as well as other types of programming that offend some persons. Copps (2003), p. 7, notes, "Compelling arguments have been made that excessive violence is every bit as indecent, profane and obscene as anything else that's broadcast."

*of being a little god, even if the great one has vanished, is the sin against the Holy Ghost.*[1]

This forceful and impressive language figures human beings as ants and grains of sand, interchangeable and uncountable. Personality is merely vanity – man who wants to believe that she is the most important theme. As for belief that persons have a unique creative power that is an irreducible part of their being, that is mocked with cutting irony as a "sin against the Holy Ghost" in a world in which clear thought and unsentimental rationality has dispelled God. A dominant tendency in the twentieth century has been to assert that human persons are more like ants than like little gods.[2] That everyone under copyright law is an author, a little god, solely by law and without the need for any distinctive acts, can be understood as legal consolation for this intellectual development.[3]

The business history of writers and photographers offers a different perspective on authors and values. Photographers' business model was much different from that of writer-authors. Photographers saw directly their end-customers and transacted directly with them   Writers, in contrast, had to make sense of many different readers. Writers confronted the risk that editors and printers would mangle the writer's work, and that publishers would capture almost all the monetary value of books, journals, magazines, and newspapers. In addition, photographers predominately generated revenue from products custom-produced to evoke a specific sense of presence. Writers generally produced a single product that readers customized in interpretation.

In occupational self-profession in the nineteenth century, photographers were not only much more numerous than writer-authors, photographers also undoubtedly earned much more from their work. A person with little education and low social status could make a large amount of money as a photographer.[4] In contrast, being an author has long been regarded as an unprofitable profession; the "Calamities of Authors," like the death of God, long proclaimed.[5] A New York journal declared in 1823:

*no encouragement whatsoever is given to the unfortunate author. The votary of the Muses, the instructor and improver of mankind, is permitted to saunter around the streets with his elbows peeping out of a more than thread-bare coat.*[6]

Without relying on their status as authors under copyright law, in the nineteenth century photographers created a much more profitable business model than writer-authors.

**III. Making a Living Creatively**

Creating new patterns of valued activity is an important aspect of creativity in work. In the U.S. in 1860, self-professed daguerreotypists and photographers were fifteen times as numerous as authors, and the former had much better earning opportunities than the latter. In 2000, photographers and "writers and authors" were about equally numerous, [1] and photographers' straight-time gross annual pay was only 58% that of writers and authors.[2] Growth in self-production of photographs contributed to the decline in the fortune of photographers. But many persons have long been able to self-produce writing.[3] Promoting progress in science and useful arts depends not just on copyright law and the economics of self-production, but also on recognizing and encouraging workers' creativity across a wide range of ordinary jobs.

---

[1] Personal letter from Oliver Wendell Holmes to Lewis Einstein, Aug. 19, 1909, printed in part in Posner (1992) pp. xxv-vi.
[2] For some outstanding and highly influential scholarly work on ants and other insects, see Wilson (1971), Hölldobler and Wilson (1994), and Eisner (2003).
[3] Distinctive acts necessary for legal recognition as an author of work under copyright, such as affixing a copyright notice in publication, were greatly lessened in 1978 and then eliminated in 1989.
[4] E.g. Bear (1873) Chap. 7. See also the description of production and pricing in Plumbe's National Gallery c. 1840 in Photographer (1896),
[5] Rice (1997) p. 94.
[6] Id., quoting the New York *Minerva*.

---

[1] Including "technical writers" with "writers and authors" would tip the balance away from photographers by about 50%. See Tables C1 and C2, above.
[2] Bureau of Labor Statistics (2002) Table 1.
[3] Tools such as word processors, the Internet, and blogs have further increased opportunities for self-production of writing. The effect of these technologies will depend greatly on the adaptability of organizations and jobs.

## Data

The first U.S. census that included detailed occupational data was in 1850. Occupation categories and counts 1850 to 1980 were taken from corresponding census report on detailed occupations. Data for 1990 are from Census (1990), Table 1. Data for 2000 are from Hecker (2001), Table 2. For discussion of revisions of the Standard Occupational Classification System, see Bureau of Labor Statistics (1999).